\documentclass[10pt,twocolumn,letterpaper]{article}

\usepackage{cvpr}
\usepackage{times}
\usepackage{epsfig}
\usepackage{graphicx}
\usepackage{amsmath}
\usepackage{amssymb}
\usepackage{framed,multirow}
\usepackage{booktabs}
\usepackage{algorithm}
\usepackage{algpseudocode}
\usepackage{setspace}

\usepackage{authblk}

\usepackage[breaklinks=true,bookmarks=false]{hyperref}

\cvprfinalcopy 

\def\cvprPaperID{****} 

\ifcvprfinal\fi
\begin{document}

\title{MetaIQA: Deep Meta-learning for No-Reference Image Quality Assessment}


\author{Hancheng~Zhu\textsuperscript{a,b}, Leida~Li\textsuperscript{a}\thanks{Corresponding author: Leida Li}~, Jinjian~Wu\textsuperscript{a}, Weisheng~Dong\textsuperscript{a}, and Guangming~Shi\textsuperscript{a}\\
\textsuperscript{a}School of Artificial Intelligence, Xidian University\\
\textsuperscript{b}School of Information and Control Engineering, China University of Mining and Technology \\
{\tt\small \textsuperscript{a}ldli@xidian.edu.cn, \textsuperscript{a}\{jinjian.wu, wsdong\}@mail.xidian.edu.cn, \textsuperscript{a}gmshi@xidian.edu.cn}  \\
{\tt\small \textsuperscript{b}zhuhancheng@cumt.edu.cn}
}

\maketitle
\thispagestyle{empty}

\begin{abstract}
Recently, increasing interest has been drawn in exploiting deep convolutional neural networks (DCNNs) for no-reference image quality assessment (NR-IQA). Despite of the notable success achieved, there is a broad consensus that training DCNNs heavily relies on massive annotated data. Unfortunately, IQA is a typical small sample problem. Therefore, most of the existing DCNN-based IQA metrics operate based on pre-trained networks. However, these pre-trained networks are not designed for IQA task, leading to generalization problem when evaluating different types of distortions. With this motivation, this paper presents a no-reference IQA metric based on deep meta-learning. The underlying idea is to learn the meta-knowledge shared by human when evaluating the quality of images with various distortions, which can then be adapted to unknown distortions easily. Specifically, we first collect a number of NR-IQA tasks for different distortions. Then meta-learning is adopted to learn the prior knowledge shared by diversified distortions. Finally, the quality prior model is fine-tuned on a target NR-IQA task for quickly obtaining the quality model. Extensive experiments demonstrate that the proposed metric outperforms the state-of-the-arts by a large margin. Furthermore, the meta-model learned from synthetic distortions can also be easily generalized to authentic distortions, which is highly desired in real-world applications of IQA metrics.
\end{abstract}


\vspace{-0.5cm}
\section{Introduction}
In recent years, the explosive growth of social networks has produced massive amounts of images. Digital images could be distorted in any stage of their life cycle, from acquisition, compression, storage to transmission, which leads to the loss of received visual information. Consequently, a reliable quality assessment metric of digital images is in great need to pick out high quality images for the end users. Although human's subjective evaluation of images is accurate and reliable, it is time-consuming and laborious in practical applications. Hence, objective image quality assessment (IQA)~\cite{Kim2017Deep} is needed to imitate human beings to automatically assess image quality, which has extensive applications in image restoration~\cite{Dong2016Image}, image retrieval~\cite{Guo2018Robust} and image quality monitoring systems~\cite{Lai2011Using}, etc.

Typically, IQA methods can be divided into three categories: full-reference IQA (FR-IQA)~\cite{KimJ2017Deep}, reduced-reference IQA (RR-IQA)~\cite{Golestaneh2016Reduced}, and no-reference IQA (NR-IQA)~\cite{Ye2012Unsupervised}, depending on the amount of reference information needed during quality evaluation. Although FR-IQA and RR-IQA methods can achieve promising performance, reference images are often not available in real-world situations. Hence, NR-IQA methods have attracted extensive attention recently, as they operate on the distorted images directly. Meantime, the lacking of reference information poses huge challenge for NR-IQA methods. Early NR-IQA methods mainly focus on specific distortion types, such as blocking artifacts~\cite{Li2014Referenceless}, blur~\cite{Li2016No-Reference} and ringing effects~\cite{Liu2010A}. The prerequisite of these approaches is that there is only one known type of distortion in the images. Since the distortion types are usually not known in advance in real-world applications, more and more attention has been drawn in general-purpose NR-IQA methods over the past few years~\cite{Saad2012Blind, Mittal2012No-Reference, Zhang2015A, Xue2013Learning, Xue2014Blind, Gu2015Using, Ye2012Unsupervised, Xu2016Blind, ghadiyaram2017perceptual}. These metrics attempt to characterize the general rules of image distortions through hand-crafted~\cite{Saad2012Blind} or learned~\cite{Ye2012Unsupervised} features, based on which an image quality prediction model can be established.

Recent years have witnessed the great success of Deep Convolutional Neural Networks (DCNNs)~\cite{He2016Deep} in many computer vision tasks~\cite{Ding2018Real-Time, Ding2019Deep}, which has also spawned a number of DCNNs-based NR-IQA approaches~\cite{Kang2014Convolutional, Bosse2018Deep, Lin2018Hallucinated, Ma2018End, Yan2019Naturalness, Zhang2019Learning, Zhang2020Blind}. These approaches have achieved significantly better performance than the traditional hand-crafted feature-based NR-IQA methods~\cite{Saad2012Blind, Mittal2012No-Reference, Zhang2015A, Xue2013Learning, Xue2014Blind, Gu2015Using}. The main reason is that DCNNs consist of massive parameters, which are helpful in learning the intricate relationship between image data and human perceived quality. At the same time, it is a broad sense that training DCNNs requires huge amount of annotated data. Unfortunately, collecting huge image quality data for training DCNNs-based IQA models is difficult, since annotating image quality by human is extremely expensive and time-consuming. As a result, the scale of existing annotated IQA databases~\cite{sheikh2005live,ponomarenko2015image} is usually limited, thus training deep IQA models directly using these databases easily leads to over-fitting. To tackle the problem, existing works usually resort to pre-trained network models where big training data is available, e.g. ImageNet image classification task~\cite{Bianco2018use, Talebi2018NIMA}. Although these metrics can alleviate the over-fitting problem to some extent, the generalization performance is unsatisfactory when facing images with unknown distortions. In our opinion, this mainly attributes to the fact that the pre-trained models are not designed for IQA task, so they cannot easily adapt to new types of distortions.

\begin{figure}[!t]
\centering
\includegraphics[width=0.37\textwidth]{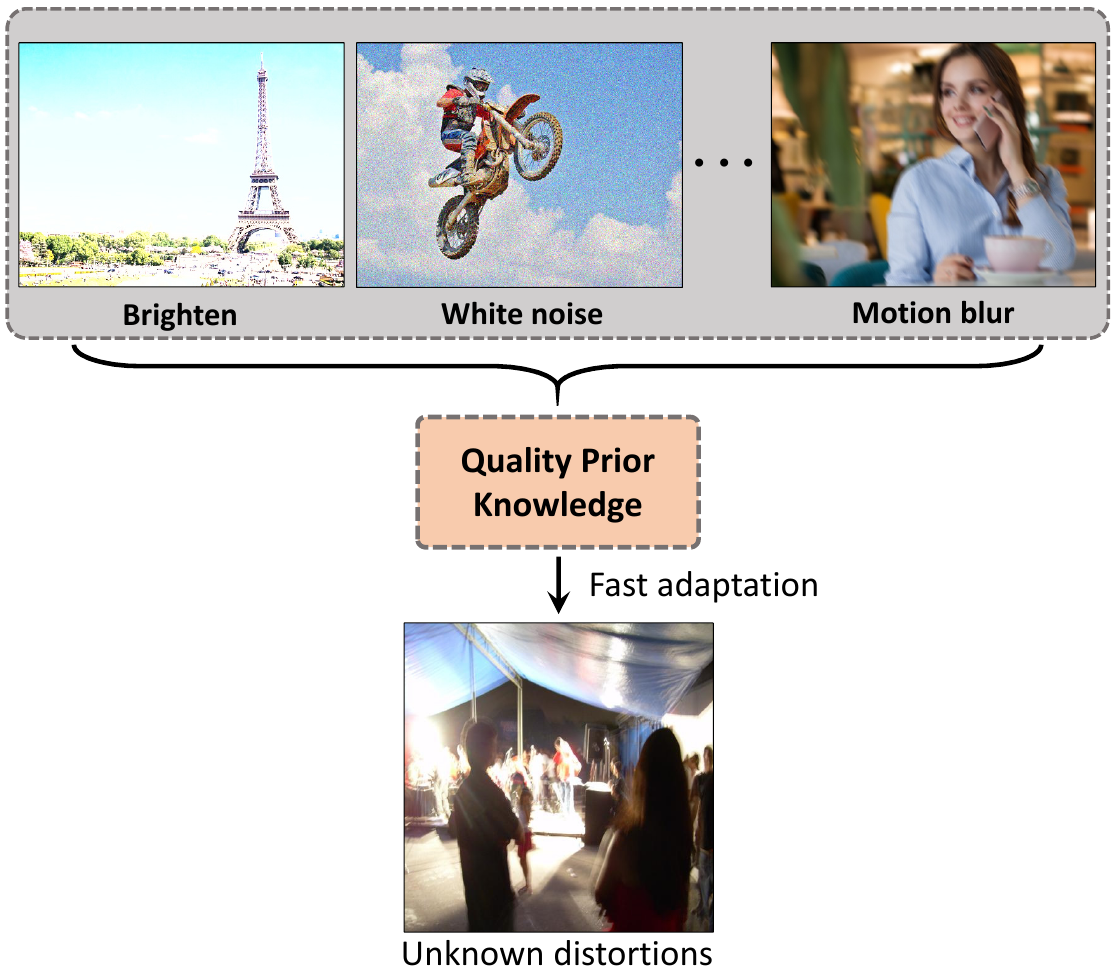}
\caption{An illustration of our motivation. Humans can use the quality prior knowledge learned from various distortions (e.g. brighten, white noise, and motion blur) for fast adapting to unknown distortions (e.g. images captured by mobile cameras). Hence, it is necessary to make the NR-IQA model learn such quality prior knowledge to achieve high generalization performance.}
\label{fig:0}
\vspace{-0.5cm}
\end{figure}

In real-world situations, human beings can easily obtain quality prior knowledge from images with various distortions and quickly adapt to the quality evaluation of unknown distorted images, as shown in Figure~\ref{fig:0}. Therefore, it is critical for NR-IQA method to learn the shared prior knowledge of humans in evaluating the quality of images with various distortions. With this motivation, this paper presents a novel NR-IQA metric based on deep meta-learning that can make machines learn to learn, that is, to have the ability to learn quickly through a relatively small amount of training samples for a related new task~\cite{Vanschoren2018Meta, Wang2019Few}. In particular, the proposed approach leverages a bi-level gradient descent strategy based on a number of distortion-specific NR-IQA tasks to learn a meta-model. The distortion-specific NR-IQA task is actually an IQA task for a specific distortion type (e.g., JPEG or blur). Different from the existing approaches, the learned meta-model can capture the shared meta-knowledge of humans when evaluating images with various distortions, enabling fast adaptation to the NR-IQA task of unknown distortions. The contributions of our work are summarized as follows.

\begin{itemize}
  \item We propose a no-reference image quality metric based on deep meta-learning. Different from the existing IQA metrics, the proposed NR-IQA model is characterized by good generalization ability, in that it can perform well on diversified distortions.
  \item We adopt meta-learning to learn the shared meta-knowledge among different types of distortions when human evaluate image quality. This is achieved using bi-level gradient optimization based on a number of distortion-specific NR-IQA tasks. The meta-knowledge serves as an ideal pre-trained model for fast adapting to unknown distortions.
  \item We have done extensive experiments on five public IQA databases containing both synthetic and authentic distortions. The results demonstrate that the proposed model significantly outperforms the state-of-the-art NR-IQA methods in terms of generalization ability and evaluation accuracy.
\end{itemize}

\section{Related Work}
\label{sec:2}
%
\subsection{No-reference image quality assessment}
NR-IQA can be classified into distortion-specific methods~\cite{Li2014Referenceless, Li2016No-Reference, Liu2010A, Wang2020Blind} and general-purpose methods~\cite{Saad2012Blind, Mittal2012No-Reference, Zhang2015A, Xue2013Learning, Xue2014Blind, Gu2015Using, Ye2012Unsupervised, Xu2016Blind, ghadiyaram2017perceptual}. In distortion-specific methods, the image quality is evaluated by extracting features of known distortion types. This kind of metrics have achieved remarkable consistency with human perception. However, their application scope is rather limited, considering the fact that the distortion type is usually unknown in real applications \cite{Jayaraman2012Objective,Ghadiyaram2016Massive}. Thus, general-purpose NR-IQA approaches have received increasingly more attention recently~\cite{Ma2019Blind}. Generally, conventional hand-crafted feature-based general-purpose NR-IQA methods can be divided into natural scene statistics (NSS) based metrics~\cite{Gao2013Universal, Mittal2012No-Reference, Moorthy2010A, Saad2012Blind} and learning-based metrics~\cite{Ye2012No-Reference, Ye2012Unsupervised, Zhang2015SOM}. The NSS-based methods hold that natural images have certain statistical characteristics, which will be changed under different distortions. Moorthy~\emph{et al.}~\cite{Moorthy2010A} proposed to extract NSS features from the discrete wavelet transform (DWT) domain for blind image quality assessment. Saad~\emph{et al.}~\cite{Saad2012Blind} leveraged the statistical features of discrete cosine transform (DCT) to estimate the image quality. Mittal~\emph{et al.}~\cite{Mittal2012No-Reference} proposed a general-purpose NR-IQA metric by extracting NSS features in the spatial domain and achieved promising performance. In additional to the NSS-based approaches, learning-based approaches have also been developed. The codebook representation approaches~\cite{Ye2012No-Reference, Ye2012Unsupervised} were proposed to predict subjective image quality scores by Support Machine Regression (SVR) model. Zhang~\emph{et al.}~\cite{Zhang2015SOM} combined the semantic-level features that influence human vision system with local features for image quality estimation.

In recent years, the deep learning-based general-purpose NR-IQA methods have demonstrated superior prediction performance over traditional methods~\cite{Bianco2018use, Talebi2018NIMA, zeng2017probabilistic, Lin2018Hallucinated, Bosse2018Deep, Ma2018End, Yan2019Naturalness, Zhang2019Learning, Zhang2020Blind}. One key issue of deep learning is that it requires abundant labeled data, but IQA is a typical small sample problem. In~\cite{Bianco2018use}, Bianco~\emph{et al.} pre-trained a deep model on the large-scale database for image classification task and then fine-tuned it for NR-IQA task. Talebi~\emph{et al.}~\cite{Talebi2018NIMA} proposed a DCNNs-based model by predicting the perceptual distribution of subjective quality opinion scores, and the model parameters were initialized by pre-training on ImageNet database~\cite{Krizhevsky2012ImageNet}. Zeng~\emph{et al.}~\cite{zeng2017probabilistic} also fine-tuned several popular pre-trained deep CNN models on IQA databases to learn a probabilistic quality representation (PQR). These approaches use the deep semantic features learned from image classification task as prior knowledge to assist in the learning of the NR-IQA task. However, image classification and quality assessment are quite different in nature, which leads to the generalization problem of deep NR-IQA models. In contrast to these approaches, in this paper, we take the advantage of meta-learning ~\cite{Vanschoren2018Meta} to explore a more effective prior knowledge for the NR-IQA task.
\subsection{Deep meta-learning}
Deep meta-learning is a knowledge-driven machine learning framework, attempting to solve the problem of how to learn~\cite{Vanschoren2018Meta}. Human beings can effectively learn a new task from limited training data, largely relying on prior knowledge of related tasks. Meta-learning is to acquire a prior knowledge model by imitating this ability of human beings. Typically, meta-learning can be divided into three main approaches: Recurrent Neural Networks (RNNs) memory-based methods~\cite{Santoro2016Meta-Learning, Munkhdalai2017Meta}, metric-based methods~\cite{Snell2017Prototypical, Sung2018Learning} and optimization-based methods~\cite{Finn2017Model, Nichol2018On}. The RNN memory-based methods use RNNs with memories to store experience knowledge from previous tasks for learning new task~\cite{Santoro2016Meta-Learning, Munkhdalai2017Meta}. The metric-based methods mainly learn an embedding function that maps the input space to a new embedding space, and leverage nearest neighbour or linear classifiers for image classification~\cite{Snell2017Prototypical, Sung2018Learning}. The optimization-based methods aim to learn the initialization parameters of a model that can quickly learn new tasks by fine-tuning the model using few training samples~\cite{Finn2017Model, Nichol2018On}. Although these methods are designed for few-shot learning in image classification task~\cite{Wang2019Few}, the optimization-based method is easier to be extended because it is based on gradient optimization without limiting network structures~\cite{Finn2017Model}. Inspired by this, we propose an optimization-based meta-learning approach for NR-IQA task, which uses a number of distortion-specific NR-IQA tasks to learn the shared prior knowledge of various distortions in images. The NR-IQA task requires a quantitative measure of the perceptual quality of image, making it more complex and difficult than image classification task. Hence, we tailor a deep meta-learning with more efficient gradient optimization.

\begin{figure*}[!t]
\centering
\includegraphics[width=0.83\textwidth]{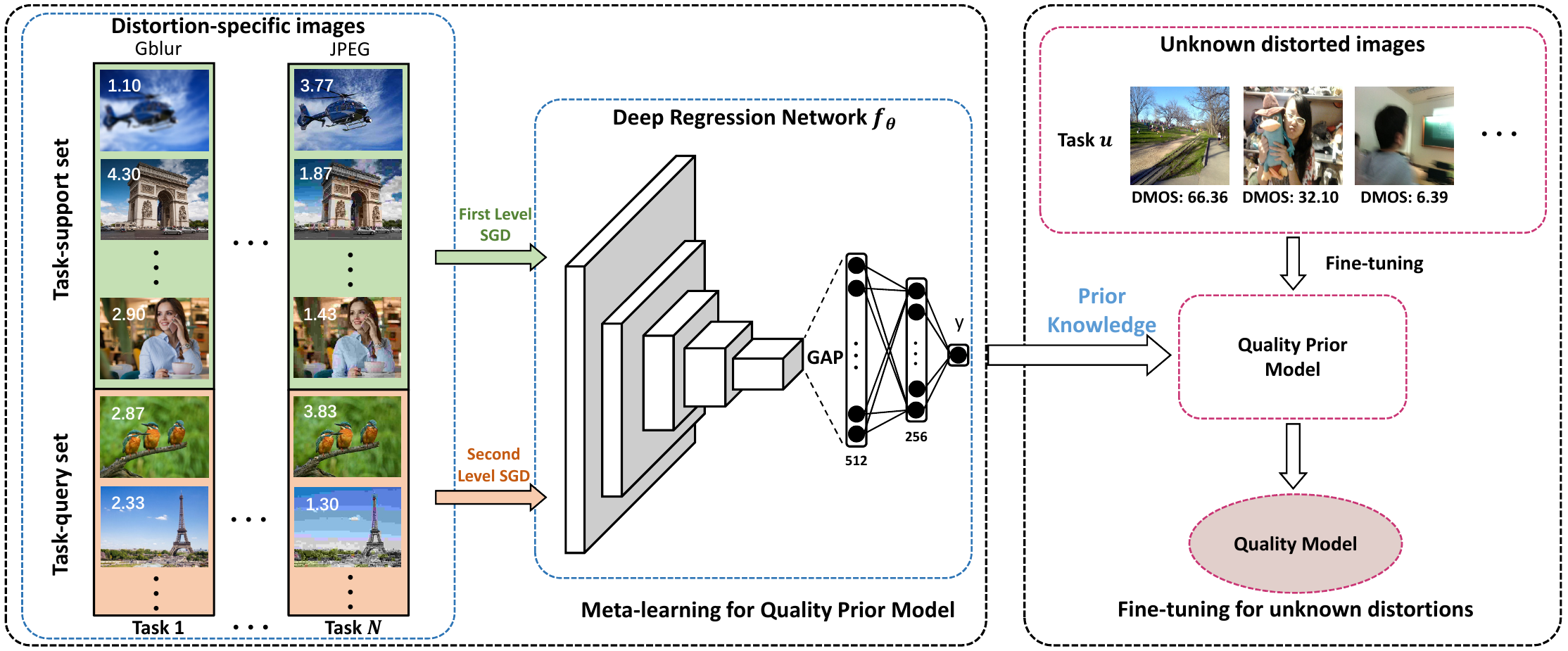}
\caption{The overview framework of our deep meta-learning approach for no-reference image quality assessment.}
\label{fig:1}
\vspace{-0.3cm}
\end{figure*}

\section{Our Approach}
\label{sec:3}
In this section, we detail our deep meta-learning approach for no-reference image quality assessment. The diversity of distortions in images leads to the generalization problem of deep NR-IQA models. In view of this, our approach leverages meta-learning to seek the general rules of image distortion through multiple distortion-specific NR-IQA tasks. That is, we learn a shared quality prior model through a number of NR-IQA tasks with known distortion types, and then fine-tune it for the NR-IQA task with unknown distortions. The overall framework of our approach is shown in Figure~\ref{fig:1}, which is composed of two steps, i.e., meta-training for quality prior model and fine-tuning for NR-IQA of unknown distortions. In the first step, we leverage a number of distortion-specific NR-IQA tasks to establish a meta-training set, which is further divided into two sets: support set and query set. Then, a bi-level gradient descent method from support set to query set is used to learn the quality prior model. In the second step, we fine-tune the quality prior model on a target NR-IQA task to obtain the quality model. Our method is termed Meta-learning based Image Quality Assessment (\textbf{MetaIQA}).

\subsection{Meta-training for quality prior model}
\textbf{Shared quality prior knowledge among distortions.} As mentioned in~\cite{Ma2019Blind}, most of the existing NR-IQA methods are distortion-aware, which are sensitive to image distortion types. Moreover, the available training data on current IQA databases cannot directly train an effective deep NR-IQA model. This limits the generalization ability of the trained NR-IQA model among images with different distortion types. Therefore, we need to learn a shared quality prior knowledge model from various distortions of images and make it quickly generalize to unknown distortions. Motivated by learning to learn in deep meta-learning~\cite{Vanschoren2018Meta}, an optimization-based approach is introduced to learn the model parameters of shared quality prior knowledge from a number of NR-IQA tasks. For the NR-IQA task, we expect that the learned model can be quickly generalized to images with unknowable distortions. Hence, we use a two-level gradient descent method to learn this generalization ability. First, the training data of each NR-IQA task is divided into support set and query set. Then, we use the support set to calculate the gradients of the model parameters and tentatively update them with stochastic gradient descent. Finally, the query set is used to verify whether the updated model is effectively performed or not. In this way, the model can learn the fast generalization ability among NR-IQA tasks with diversified distortions. The two-level gradient descent approach from support set to query set is called bi-level gradient optimization.

\textbf{Meta-learning with bi-level gradient optimization.} Since the optimization-based meta-learning method can be easily applied to any deep network using stochastic gradient descent, we introduce a deep regression network $f_{\theta}$ for the NR-IQA task. As shown in Figure~\ref{fig:1}, the deep regression network consists of convolutional layers and fully-connected layers. The convolutional layers derive from a popular deep network and we employ a Global Average Pooling (GAP) operation for yielding a fully-connected layer. Then, we add another fully-connected layer to generate the output of our deep regression network. In particular, for an input image $x$, we fed it into the deep network to generate the predicted quality score of the image $\hat{y}$, which is defined as
\begin{equation}
\label{eq1}
 \hat{y} = f_{\theta} (x; \theta),
\end{equation}
where $\theta$ denotes the initialized network parameters. Since we expect to minimize the difference between the predicted and ground-truth quality scores of the image $x$, the squared Euclidean distance is used as loss function, which takes the following form
\begin{equation}
\label{eq2}
  \mathcal{L} = \Vert f_{\theta} (x; \theta) - y \Vert_{2}^{2},
\end{equation}
where $y$ denotes the ground-truth quality score of the input image $x$.

The purpose of our approach is to learn a shared prior model among various distortions when human evaluate image quality. Therefore, we obtain the meta-training set $\mathcal{D}_{meta}^{p(\tau)} = \{\mathcal{D}_{s}^{\tau_{n}}, \mathcal{D}_{q}^{\tau_{n}}\}_{n=1}^{N}$ from a number of distortion-specific NR-IQA tasks, where $\mathcal{D}_{q}^{\tau_{n}}$ and $\mathcal{D}_{s}^{\tau_{n}}$ are the support set and query set of each task, and $N$ is the total number of tasks. In order to capture a generalized model among different NR-IQA tasks, we randomly sample $k$ tasks as a mini-batch from the meta-training set ($1<k\leq N$). For the $i$-th support set $\mathcal{D}_{s}^{\tau_{i}}$ in the mini-batch, the loss can be calculated by Eq.~\ref{eq2} and denoted as $\mathcal{L}_{\tau_{i}}$ $(i \in \{1,2,\ldots,k\})$. Since our deep regression network is more complex than the classification network in~\cite{Finn2017Model} and there are more samples available for training, we leverage a more efficient stochastic gradient descent (SGD) approach to optimize our model. Therefore, we first calculate the first-order gradients of loss function $\mathcal{L}_{\tau_{i}}$ relating to all model parameters and it is defined as
\begin{equation}
\label{eq3}
  g_{\theta} = \nabla_{\theta} \mathcal{L}_{\tau_{i}}(f_{\theta}).
\end{equation}
Then, we update the model parameters for $S$ steps using the Adam~\cite{Kingma2015Adam} optimizer on the support set $\mathcal{D}_{s}^{\tau_{i}} (i=1,2,...,k)$, which is defined as
\begin{equation}
\label{eq4}
  Adam(\mathcal{L}_{\tau_{i}}, \theta): \theta_i^{'} \leftarrow \theta - \alpha \sum_{s=1}^S \frac{m_{\theta^{(s)}}}{\sqrt{v_{\theta^{(s)}}}+\epsilon},
\end{equation}
where $\epsilon = 1e-8$ and $\alpha$ is the inner learning rate. $m_{\theta^{(s)}}$ and $v_{\theta^{(s)}}$ denote the first moment and second raw moment of gradients, which are formulated as
\begin{equation}
\label{eq5}
  m_{\theta^{(s)}} =  \mu_1 m_{\theta^{(s-1)}} + (1-\mu_1)g_{\theta^{(s)}},
\end{equation}
\begin{equation}
\label{eq6}
  v_{\theta^{(s)}} = \mu_2 v_{\theta^{(s-1)}} + (1-\mu_2) g_{\theta^{(s)}}^2,
\end{equation}
where $m_{\theta^{(0)}}=0$ and $v_{\theta^{(0)}}=0$. $\mu_1$ and $\mu_2$ are the exponential decay rates of $m_{\theta^{(s)}}$ and $v_{\theta^{(s)}}$, respectively. $g_{\theta^{(s)}}$ denote the updated gradients in step $s$ $(s \in \{1,2,...,S\})$. As we mentioned previously, we expect that the quality model updated with the support set can perform well on the query set. In contrast to calculating second-order gradients in~\cite{Finn2017Model}, we then compute the first-order gradients of updated model parameters for a second time to reduce the computational complexity of our model. The model parameters $\theta_i^{'}$ are updated with Adam optimizer for $S$ steps on the query set $\mathcal{D}_{q}^{\tau_{i}} (i=1,2,...,k)$, which takes the following form
\begin{equation}
\label{eq7}
  Adam(\mathcal{L}_{\tau_{i}}, \theta_i^{'}): \theta_i \leftarrow \theta_i^{'} - \alpha \sum_{s=1}^S \frac{m_{\theta^{'(s)}}}{\sqrt{v_{\theta^{'(s)}}}+\epsilon},
\end{equation}
where $m_{\theta^{'(s)}}$ and $v_{\theta^{'(s)}}$ are the first moment and second raw moment of gradients. For the mini-batch of $k$ tasks, the gradients of all tasks are integrated to update the final model parameters, which is defined as
\begin{equation}
\label{eq8}
\theta \leftarrow \theta - \beta \frac{1}{k}\sum_{i=1}^k( \theta - \theta_i),
\end{equation}
where $\beta$ is the outer learning rate. With this approach, we iteratively sample $k$ NR-IQA tasks on the meta-training set $\mathcal{D}_{meta}^{p(\tau)}$ to train our deep regression network $f_{\theta}$. Finally, the quality prior model shared with various image distortions can be obtained by the meta-learning with bi-level gradient optimization.
\subsection{Fine-tuning for unknown distortions}
After training the quality prior model from a number of distortion-specific NR-IQA tasks, we then use this model as prior knowledge for fine-tuning on NR-IQA task with unknown distortions. Given $M$ training images with annotated quality scores from a target NR-IQA task, we denote the predicted and ground-truth quality scores of $i$-th image as $\hat{y}_{i}$ and $y_{i}$ $(i = 1,2,...,M)$, respectively. We first use the squared Euclidean distance as loss function, which is formulated as
\begin{equation}
\label{eq9}
  \mathcal{L} = \frac{1}{M}\sum_{i=1}^{M}\Vert \hat{y}_i - y_{i} \Vert_{2}^{2}.
\end{equation}
Then, we leverage Adam optimizer to update the quality prior model for $P$ steps on the NR-IQA task and it is defined as
\begin{equation}
\label{eq10}
  Adam(\mathcal{L}, \theta): \theta_{te} \leftarrow \theta - \alpha_f \sum_{p=1}^P \frac{m_{\theta^{(p)}}}{\sqrt{v_{\theta^{(p)}}}+\epsilon},
\end{equation}
where $\alpha_f$ is the learning rate of fine-tuning. $m_{\theta^{(p)}}$ and $v_{\theta^{(p)}}$ are first moment and second raw moment of gradients. Finally, the quality model can be obtained for assessing the quality of images with unknown distortions. It is worth noting that the fine-tuning process of our approach does not need to learn additional parameters, which greatly improves the learning efficiency and enhances the generalization ability of our model.

For a query image $x$, the predicted quality score $\hat{y}$ can be obtained by capturing the output of the quality model $\hat{y}=f_{\theta_{te}} (x; \theta_{te})$. The whole procedure of the proposed MetaIQA is summarized in Algorithm~\ref{alg:Framwork}.

\begin{algorithm}[htb]
\setstretch{1}
\caption{ Meta-learning based IQA (MetaIQA)}
\label{alg:Framwork}
\begin{algorithmic}[1]
\Require
Meta-training set $\mathcal{D}_{meta}^{p(\tau)} = \{\mathcal{D}_{s}^{\tau_{i}}, \mathcal{D}_{q}^{\tau_{i}}\}_{i=1}^{N}$, where $\mathcal{D}_{tr_q}^{\tau_{i}}$ and $\mathcal{D}_{tr_s}^{\tau_{i}}$ are task-support set and task-query set, and $N$ is the total number of tasks, a target NR-IQA task with $M$ training images, query image $x$, learning rate $\beta$
\Ensure
Predicted quality score $\hat{y}$ for $x$
\State Initialize model parameters $\theta$; \\
 {$/\star$ \verb"meta-training for prior model"  $\star/$}
\For{ \textit{iteration} $= 1, 2, ...$}
\State Sample a mini-batch of $k$ tasks in $\mathcal{D}_{meta}^{p(\tau)}$;
\For{$i=1, 2, ..., k$}  \\
\qquad \quad $/\star$ \verb"first level computing"  $\star/$
\State Compute $\theta_i^{'} = Adam(\mathcal{L}_{\tau_i}, \theta)$ on $\mathcal{D}_{s}^{\tau_{i}}$; \\
\qquad \quad $/\star$ \verb"second level computing"  $\star/$
\State Compute $\theta_i = Adam(\mathcal{L}_{\tau_i}, \theta_i^{'})$ on $\mathcal{D}_{q}^{\tau_{i}}$;
\EndFor
\State update $\theta \leftarrow \theta - \beta \frac{1}{k} \sum_{i=1}^k( \theta - \theta_i)$;
\EndFor \\
$/\star$ \verb"fine-tuning for NR-IQA task"  $\star/$
\State Update $\theta_{te} = Adam(\mathcal{L}, \theta)$ on the NR-IQA task;
\State Input $x$ into the quality model $f_{\theta_{te}}$; \\
\Return $\hat{y}$.
\end{algorithmic}
\end{algorithm}

\vspace{-0.3cm}
\section{Experiments}
\label{sec:4}
\subsection{Databases}
We evaluate the performance of our approach on two kind of databases: synthetically distorted IQA databases and authentically distorted IQA databases.

\textbf{Synthetically distorted IQA databases} can be used for generating the meta-training set and evaluating the generalization performance of our quality prior model for unseen distortions, including TID2013~\cite{ponomarenko2015image} and KADID-10K~\cite{lin2019kadid}. The information for each database is listed in Table \ref{table1}. 

\textbf{Authentically distorted IQA databases} are used to verify the generalization performance of our quality prior model for real distorted images, including CID2013~\cite{Virtanen2015CID2013}, LIVE challenge~\cite{Ghadiyaram2016Massive} and KonIQ-10K~\cite{lin2018koniq}. The CID2013 database contains six subsets with a total of 480 authentically distorted images captured by 79 different digital cameras. Subjects participated in the user study to evaluate the quality scores of images, which are in the range $[0,~100]$, and the higher the score, the higher the quality. The LIVE challenge database contains 1,162 images with authentic distortions taken from mobile cameras, such as motion blur, overexposure or underexposure, noise and JPEG compression. The quality scores of images were obtained by crowdsourcing experiments, which are in the range $[0,~100]$, and higher score indicates higher quality. Recently, a relatively large-scale IQA database, KonIQ-10K, consisting of 10,073 images was introduced in~\cite{lin2018koniq}. The quality score of each image is averaged by the five-point ratings of about 120 workers, which are in the range $[1,~5]$, and higher score also indicates higher quality.

\begin{table}
\caption{Summary of synthetically distorted IQA databases with respect to numbers of reference images (Ref.), distortion images (Dist.), distortion types (Dist. Types) and score range. higher score indicates higher quality.}
\fontsize{7pt}{7pt}\selectfont%
\renewcommand{\arraystretch}{1.1}
\begin{center}
\label{table1}
\begin{tabular}{cccccc}
\toprule[1pt]
Databases &Ref. &Dist.   &Dist. Types &Score Range \\
\noalign{\smallskip}\hline\noalign{\smallskip}
TID2013~\cite{ponomarenko2015image}    &25  &3,000 &24  &[0,~9]\\
KADID-10K~\cite{lin2019kadid}    &81  &10,125 &25 &[1,~5]\\
\bottomrule[1.2pt]
\end{tabular}
\end{center}
\vspace{-0.7cm}
\end{table}

\subsection{Implementation details}
In the proposed model, a popular network architecture, ResNet18~\cite{He2016Deep}, is adopted as our backbone network. All training images are randomly cropped to $224\times224$ pixel patches for feeding into the proposed model. We train our model using bi-level gradient optimization with the inner learning rate $\alpha$ of $1e-4$ and the outer learning rate $\beta$ of $1e-2$, which is implemented based on Pytorch~\cite{paszke2017automatic}. We set the fine-tuning learning rate $\alpha_f$ to $1e-5$. These learning rates drop to a factor of 0.8 after every five epochs and the total epoch is 100. For both model training and fine-tuning, the weight decay is $1e-5$. The other hyper-parameters are set as follows: mini-batch size $k$ of 5, exponential decay rate $\mu_1$ of 0.9, exponential decay rate $\mu_2$ of 0.99, learning steps $S$ of 6, learning steps $P$ of 15.

\subsection{Evaluation criteria}
In our experiments, Spearman's rank order correlation coefficient (SROCC) and Pearson's linear correlation coefficient (PLCC) are employed to evaluate the performance of NR-IQA methods~\cite{Bosse2018Deep, Yan2019Naturalness}. For $N$ testing images, the PLCC is defined as
\begin{equation}
\label{eq11}
  \textrm{PLCC} = \frac{\sum_{i=1}^{N}(s_i - \mu_{s_i})(\hat{s}_i - \mu_{\hat{s}_i})}{\sqrt{\sum_{i=1}^{N}(s_i - \mu_{s_i})^2}\sqrt{\sum_{i=1}^{N}(\hat{s}_i - \mu_{\hat{s}_i})^2}},
\end{equation}
where $s_i$ and $\hat{s}_i$ denote the ground-truth and predicted quality scores of $i$-th image, and $\mu_{s_i}$ and $\mu_{\hat{s}_i}$ denote the average of each. Let $d_i$ denote the difference between the ranks of $i$-th test image in ground-truth and predicted quality scores, the SROCC is defined as
\begin{equation}
\label{eq12}
  \textrm{SROCC} = 1 - \frac{6\sum_{i=1}^{N}d_i^2}{N(N^2-1)}.
\end{equation}
The PLCC and SROCC range from -1 to 1, and higher absolute value indicates better prediction performance.

\subsection{Comparisons with the state-of-the-arts}
\textbf{Evaluation on synthetically distorted images.} To validate the generalization performance of our meta-model for unknown distortions, we compare our method with six state-of-the-art general-purpose NR-IQA methods by using the Leave-One-Distortion-Out cross validation on TID2013~\cite{ponomarenko2015image} and KADID-10K~\cite{lin2019kadid} databases. In implementation, suppose there are
$N$ kinds of distortions in a database, we use ($N - 1$) kinds of distortions for training and the remaining one kind of distortion is used for performance test. These methods are BLIINDS-II~\cite{Saad2012Blind}, BRISQUE~\cite{Mittal2012No-Reference}, ILNIQE~\cite{Zhang2015A}, CORNIA~\cite{Ye2012Unsupervised}, HOSA~\cite{Xu2016Blind} and WaDIQaM-NR~\cite{Bosse2018Deep}. For a fair comparison, all the source codes of NR-IQA methods released by original authors are conducted under the same training-testing strategy.

The tested SROCC values of our approach and state-of-the-art NR-IQA methods are listed in Table \ref{table2} and the best result for each distortion type is marked in bold. As can be seen, our approach is superior to other methods in overall performance (average results) on both databases by a large margin. For most of the distortion types (19 out of 24 on TID2013 and 19 out of 25 on KADID-10K), our method can achieve the best evaluation performance. In TID2013 database, the SROCC values of our method are higher than 0.9 for more than half of the distortion types, which indicates that our meta-learning based NR-IQA method can effectively learn a shared quality prior model and fast adapt to a NR-IQA task with unknown distortion types.

\begin{table*}
\caption{SROCC values comparison in leave-one-distortion-out cross validation on TID2013 and KADID-10K databases.}
\fontsize{7pt}{7pt}\selectfont%
\renewcommand{\arraystretch}{1}
\begin{center}
\vspace{0.1cm}
\label{table2}
\begin{tabular}{c|c|ccccc|cc}
\toprule[1pt]
\multirow{25}{13mm}{TID2013}
&Dist. type &BLIINDS-II~\cite{Saad2012Blind} &BRISQUE~\cite{Mittal2012No-Reference} &ILNIQE~\cite{Zhang2015A}  &CORNIA~\cite{Ye2012Unsupervised} &HOSA~\cite{Xu2016Blind} &WaDIQaM-NR~\cite{Bosse2018Deep} &\textbf{MetaIQA}  \\
\hline\noalign{\smallskip}
&AGN    &0.7984  &0.9356 &0.8760 &0.4465 &0.7582 &0.9080  &\textbf{0.9473} \\
&ANC &0.8454  &0.8114 &0.8159 &0.1020 &0.4670 &0.8700 &\textbf{0.9240} \\
&SCN    &0.6477  &0.5457 &0.9233 &0.6697 &0.6246 &0.8802 &\textbf{0.9534} \\
&MN    &0.2045  &0.5852 &0.5120 &0.6096 &0.5125 &\textbf{0.8065} &0.7277 \\
&HFN    &0.7590  &0.8965 &0.8685 &0.8402 &0.8285 &0.9314 &\textbf{0.9518} \\
&IN  &0.5061  &0.6559 &0.7551 &0.3526 &0.1889 &\textbf{0.8779} &0.8653 \\
&QN &0.3086 &0.6555 &\textbf{0.8730} &0.3723 &0.4145 &0.8541 &0.7454  \\
&GB   &0.9069  &0.8656 &0.8142 &0.8879 &0.7823 &0.7520  &\textbf{0.9767} \\

&DEN &0.7642  &0.6143 &0.7500 &0.6475 &0.5436 &0.7680  &\textbf{0.9383} \\
&JPEG &0.7951  &0.5186 &0.8349 &0.8295 &0.8318 &0.7841  &\textbf{0.9340} \\
&JP2K  &0.8221  &0.7592 &0.8578 &0.8611 &0.5097 &0.8706  &\textbf{0.9586} \\
&JGTE  &0.4509  &0.5604 &0.2827 &0.7282 &0.4494 &0.5191  &\textbf{0.9297} \\
&J2TE &0.7281  &0.7003 &0.5248 &0.4817 &0.1405 &0.4322  &\textbf{0.9034} \\
&NEPN    &0.1219  &0.3111 &-0.0805 &0.3571 &0.2163 &0.1230  &\textbf{0.7238} \\
&Block &0.2789  &0.2659 &-0.1357 &0.2345 &0.3767 &\textbf{0.4059}  &0.3899 \\
&MS    &0.0970  &0.1852 &0.1845 &0.1775 &0.0633 &\textbf{0.4596} &0.4016 \\

&CTC    &0.3125  &0.0182 &0.0141 &0.2122 &0.0466 &0.5401 &\textbf{0.7637} \\
&CCS    &0.0480  &0.2142 &-0.1628 &0.2299 &-0.1390 &0.5640 &\textbf{0.8294} \\
&MGN  &0.7641  &0.8777 &0.6932 &0.4931 &0.5491 &0.8810  &\textbf{0.9392} \\
&CN &0.0870  &0.4706 &0.3599 &0.5069 &0.3740 &0.6466  &\textbf{0.9516}\\
&LCNI  &0.4480  &0.8238 &0.8287 &0.7191 &0.5053 &0.6882  &\textbf{0.9779} \\
&ICQD &0.7953  &0.4883 &0.7487 &0.7757 &0.8036 &0.7965  &\textbf{0.8597} \\
&CHA &0.5417  &0.7470 &0.6793 &0.6937 &0.6657 &0.7950  &\textbf{0.9269} \\
&SSR   &0.7416  &0.7727 &0.8650 &0.8867 &0.8273 &0.8220  &\textbf{0.9744}\\
\cline{2-9}\noalign{\smallskip}
&Average   &0.5322 &0.5950 &0.5701 &0.5465 &0.4725 &0.7073  &\textbf{0.8539} \\
\bottomrule[1pt]\noalign{\smallskip}
\multirow{26}{13mm}{KADID-10K}
&GB    &0.8799  &0.8118 &0.8831 &0.8655 &0.8522 &0.8792  &\textbf{0.9461} \\
&LB &0.7810  &0.6738 &0.8459 &0.8109 &0.7152 &0.7299 &\textbf{0.9168} \\
&MB    &0.4816  &0.4226 &0.7794 &0.5323 &0.6515 &0.7304 &\textbf{0.9262} \\
&CD    &0.5719  &0.5440 &0.6780 &0.2432 &0.7272 &0.8325 &\textbf{0.8917} \\
&CS    &-0.1392  &-0.1821 &0.0898 &-0.0023 &0.0495 &0.4209 &\textbf{0.7850} \\
&CQ  &0.6695  &0.6670 &0.6763 &0.3226 &0.6617 &\textbf{0.8055} &0.7170 \\
&CSA1 &0.0906  &0.0706 &0.0266 &-0.0194 &0.2158 &0.1479 &\textbf{0.3039}  \\
&CSA2  &0.6017  &0.3746 &0.6771 &0.1197 &0.8408 &0.8358 &\textbf{0.9310} \\

&JP2K &0.6546  &0.5159 &0.7895 &0.3417 &0.6078 &0.5387 &\textbf{0.9452} \\
&JPEG &0.4140  &0.7821 &0.8036 &0.5561 &0.5823 &0.5298 &\textbf{0.9115} \\
&WN   &0.6277  &0.7080 &0.7757 &0.3574 &0.6796 &0.8966 &\textbf{0.9047} \\
&WNCC  &0.7567  &0.7182 &0.8409 &0.4183 &0.7445 &0.9247 &\textbf{0.9303} \\
&IN &0.5469  &-0.5425 &0.8082 &0.2188 &0.2535 &0.8142 &\textbf{0.8673} \\
&MN    &0.7017  &0.6741 &0.6824 &0.3060 &0.7757 &0.8841 &\textbf{0.9247} \\
&Denoise &0.4566  &0.2213 &0.8562 &0.2293 &0.2466 &0.7648 &\textbf{0.8985} \\
&Brighten    &0.4583  &0.5754 &0.3008 &0.2272 &0.7525 &0.6845 &\textbf{0.7827} \\
&Darken    &0.4391  &0.4050 &0.4363 &0.2060 &\textbf{0.7436} &0.2715 &0.6219 \\

&MS    &0.1119  &0.1441 &0.3150 &0.1215 &\textbf{0.5907} &0.3475 &0.5555 \\
&Jitter  &0.6287  &0.6719 &0.4412 &0.7186 &0.3907 &0.7781 &\textbf{0.9278} \\
&NEP &0.0832  &0.1911 &0.2178 &0.1206 &\textbf{0.4607} &0.3478 &0.4184 \\
&Pixelate   &0.1956  &0.6477 &0.5770 &0.5868 &0.7021 &0.6998 &\textbf{0.8090} \\
&Quantization &0.7812  &0.7135 &0.5714 &0.2592 &0.6811 &0.7345 &\textbf{0.8770} \\
&CB &-0.0204  &0.0673 &0.0029 &0.0937 &0.3879 &0.1602 &\textbf{0.5132} \\
&HS   &-0.0151  &0.3611 &\textbf{0.6809} &0.1142 &0.2302 &0.5581 &0.4374 \\
&CC   &0.0616  &0.1048 &0.0723 &0.1253 &\textbf{0.4521} &0.4214 &0.4377 \\
\hline\noalign{\smallskip}
&Average   &0.4328  &0.4136 &0.5528 &0.3149 &0.5598 &0.6295 &\textbf{0.7672} \\
\bottomrule[1pt]
\end{tabular}
\end{center}
\vspace{-0.5cm}
\end{table*}

\textbf{Generalization performance on authentically distorted images.} To further evaluate the generalization performance of the quality prior model learned from synthetic distortions for the IQA of authentic distortions, we compare the proposed method with five state-of-the-art hand-crafted feature-based and six state-of-the-art deep learning-based general-purpose NR-IQA methods. The five hand-crafted feature-based NR-IQA methods are BLIINDS-II~\cite{Saad2012Blind}, BRISQUE~\cite{Mittal2012No-Reference}, ILNIQE~\cite{Zhang2015A}, CORNIA~\cite{Ye2012Unsupervised} and HOSA~\cite{Xu2016Blind}, while the six deep learning-based NR-IQA methods are BIECON~\cite{Kim2017Fully}, MEON~\cite{Ma2018End}, WaDIQaM-NR~\cite{Bosse2018Deep}, DistNet-Q3~\cite{Dendi2019Generating}, DIQA~\cite{Kim2019DeepCNN} and NSSADNN~\cite{Yan2019Naturalness}. For a fair comparison with the reported results of these methods on CID2013~\cite{Virtanen2015CID2013}, LIVE challenge~\cite{Ghadiyaram2016Massive} and KonIQ-10K~\cite{lin2018koniq} databases, we follow the same experimental setup in~\cite{Bosse2018Deep, Xu2016Blind, Yan2019Naturalness}. In CID2013 database, four out of six subsets are used for model training, and the remaining two subsets are used for testing. In LIVE challenge and KonIQ-10K databases, all images are randomly divided into 80\% training samples and 20\% testing samples. All experiments are conducted 10 times to avoid the bias of randomness and the average results of PLCC and SROCC are reported.

In our approach, we first normalize the subjective scores of images on TID2013 and KADID-10K databases to $[0,~1]$ and then use the generated NR-IQA tasks to train our network for obtaining a quality prior model. Finally, we fine-tune the quality prior model on the training set of CID2013, LIVE challenge and KonIQ-10K. Table \ref{table3} summarizes the testing results on the three IQA databases and the best results among the NR-IQA methods for each database are shown boldfaced. We can see that our approach achieves the best evaluation performance on LIVE challenge and KonIQ-10K databases. Our method and NSSADNN have achieved comparable results on CID2013 database, which are significantly better than other NR-IQA methods. This indicates that our method based on meta-learning can capture the quality prior model shared by human when evaluating the perceived quality images with various synthesized distortions, and then quickly adapt to a NR-IQA task with authentic distortions.

\subsection{Visual analysis for quality prior model}
\begin{figure}[!t]
\centering
\includegraphics[width=0.49\textwidth]{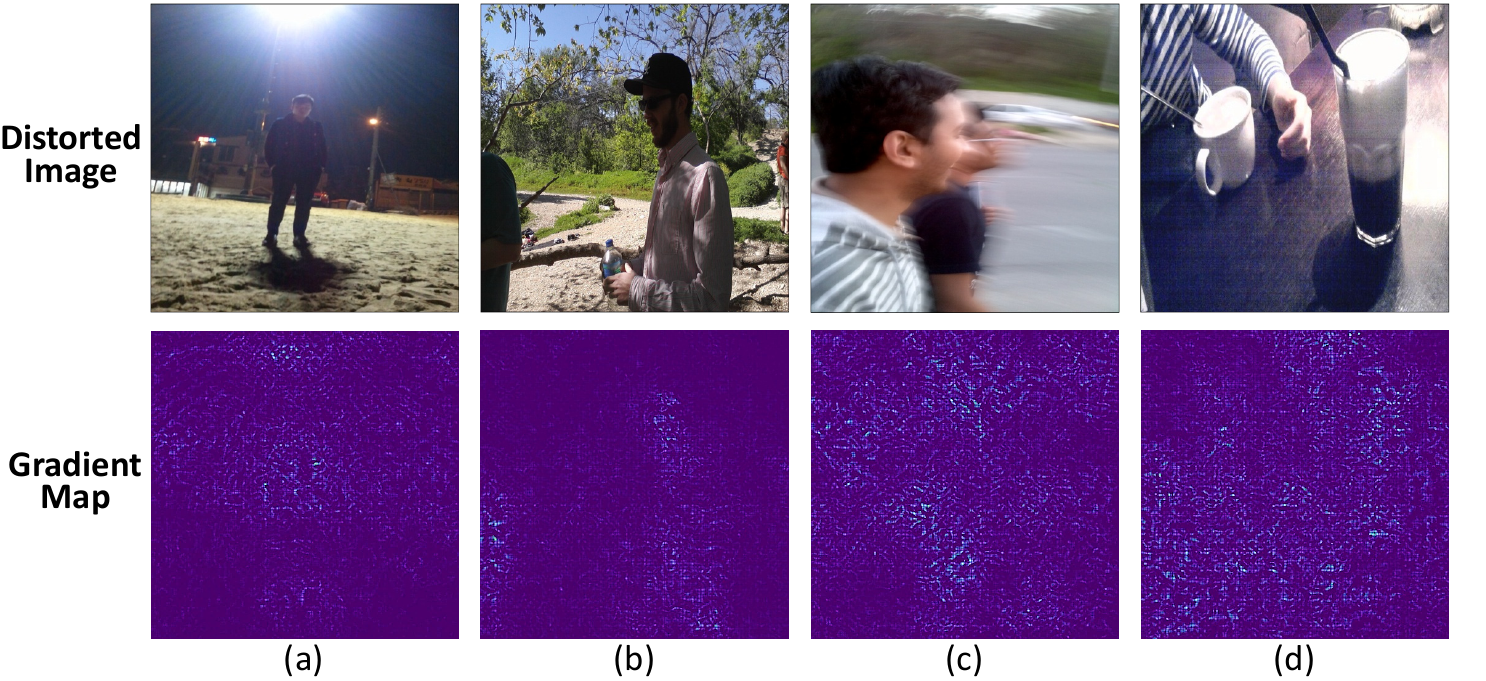}
\caption{The gradient maps of some authentically distorted images in LIVE challenge database.}
\label{fig:3}
\end{figure}

In this section, we performed a visual experiment to demonstrate the effectiveness of our quality prior model. Particularly, we use a CNN visualization code\footnote{https://github.com/sar-gupta/convisualize\_nb} to show the gradient maps in pixel-wisely with various distortions. We learn the quality prior model from distortion-specific images on TID2013 and KADID-10K databases, and then randomly select four severely distorted images in the LIVE challenge database for visualization experiment. The images as well as the corresponding gradient maps are shown in Figure~\ref{fig:3}. As can be seen, the gradient maps can accurately capture the location of authentic distortions in images, such as overexposure in Figure~\ref{fig:3}(a), underexposure in Figure~\ref{fig:3}(b), motion blur in Figure~\ref{fig:3}(c) and noise in Figure~\ref{fig:3}(d). This strongly demonstrates that the shared prior knowledge of various distortions in images can be effectively learned from a number of NR-IQA tasks through meta-learning.

\begin{table}
\caption{Comparison results (PLCC and SROCC) of our approach with several state-of-the-art NR-IQA methods on authentically distorted IQA databases (i.e., CID2013~\cite{Virtanen2015CID2013}, LIVE challenge~\cite{Ghadiyaram2016Massive} and KonIQ-10K~\cite{lin2018koniq}).}
\fontsize{6pt}{6pt}\selectfont%
\renewcommand{\arraystretch}{1}
\begin{center}
\label{table3}
\begin{tabular}{l|c|c|c|c|c|c}
\toprule[1pt]
\multirow{2}{1mm}{Methods}
&\multicolumn{2}{c|}{CID2013} &\multicolumn{2}{c|}{LIVE challenge}  &\multicolumn{2}{c}{KonIQ-10K}\\
\cline{2-7}\noalign{\smallskip}
&PLCC &SROCC &PLCC &SROCC &PLCC &SROCC \\
\hline\noalign{\smallskip}
BLIINDS-II~\cite{Saad2012Blind}  &0.565 &0.487  &0.507 &0.463 &0.615 &0.529\\
BRISQUE~\cite{Mittal2012No-Reference} &0.648 &0.615  &0.645 &0.607 &0.537 &0.473\\
ILNIQE~\cite{Zhang2015A}  &0.538 &0.346   &0.589 &0.594 &0.537 &0.501\\
CORNIA~\cite{Ye2012Unsupervised}  &0.680 &0.624   &0.662 &0.618 &0.795 &0.780\\
HOSA~\cite{Xu2016Blind}   &0.685 &0.663  &0.678 &0.659 &0.813 &0.805\\
\hline\noalign{\smallskip}
BIECON~\cite{Kim2017Fully} &0.620 &0.606 &0.613 &0.595 &/ &/\\
MEON~\cite{Ma2018End} &0.703 &0.701  &0.693 &0.688 &/ &/\\
WaDIQaM-NR~\cite{Bosse2018Deep} &0.729 &0.708 &0.680 &0.671 &0.761 &0.739\\
DistNet-Q3~\cite{Dendi2019Generating} &/ &/ &0.601 &0.570 &0.710 &0.702 \\
DIQA~\cite{Kim2019DeepCNN}  &0.720 &0.708  &0.704 &0.703 &/ &/\\
NSSADNN~\cite{Yan2019Naturalness} &\textbf{0.825} &0.748 &0.813 &0.745 &/ &/\\
\textbf{MetaIQA} &0.784 &\textbf{0.766} &\textbf{0.835} &\textbf{0.802} &\textbf{0.887} &\textbf{0.850}\\
\bottomrule[1pt]
\end{tabular}
\end{center}
\vspace{-0.5cm}
\end{table}

\subsection{Ablation study}
To further investigate whether the effectiveness of our approach is derived from meta-learning, we conduct ablation studies in this experiment. The baseline method is to first train our network model by directly using the Adam optimizer on distortion-specific images, and then fine-tune the model on the training set of authentically distorted images (called Baseline). It is worth noting that baseline method and our method have the same number of network parameters but are trained by two different optimization approaches. The results of all tested images on three authentically distorted IQA databases are summarized in Table \ref{table4}. From the results, we can see that our MetaIQA method is superior to Baseline method by a large margin on all databases. Compared with the baseline approach, MetaIQA has better generalization performance and can improve the performance of NR-IQA model without changing the network structure. This demonstrates the effectiveness of our method in dealing with the NR-IQA task.

\begin{table}
\caption{Ablation study results (PLCC and SROCC) on authentically distorted IQA databases (i.e., CID2013~\cite{Virtanen2015CID2013}, LIVE challenge~\cite{Ghadiyaram2016Massive} and KonIQ-10K~\cite{lin2018koniq}).}
\fontsize{7pt}{7pt}\selectfont%
\renewcommand{\arraystretch}{1}
\begin{center}
\label{table4}
\begin{tabular}{l|c|c|c|c|c|c}
\toprule[1pt]
\multirow{2}{1mm}{Methods}
&\multicolumn{2}{c|}{CID2013} &\multicolumn{2}{c|}{LIVE challenge}  &\multicolumn{2}{c}{KonIQ-10K}\\
\cline{2-7}\noalign{\smallskip}
&PLCC &SROCC &PLCC &SROCC &PLCC &SROCC \\
\hline\noalign{\smallskip}
Baseline  &0.727 &0.712  &0.801 &0.743 &0.832 &0.816\\
MetaIQA &\textbf{0.784} &\textbf{0.766} &\textbf{0.835} &\textbf{0.802} &\textbf{0.887} &\textbf{0.850}\\
\bottomrule[1pt]
\end{tabular}
\end{center}
\vspace{-0.5cm}
\end{table}

\begin{figure}
\centering
\includegraphics[width=0.43\textwidth]{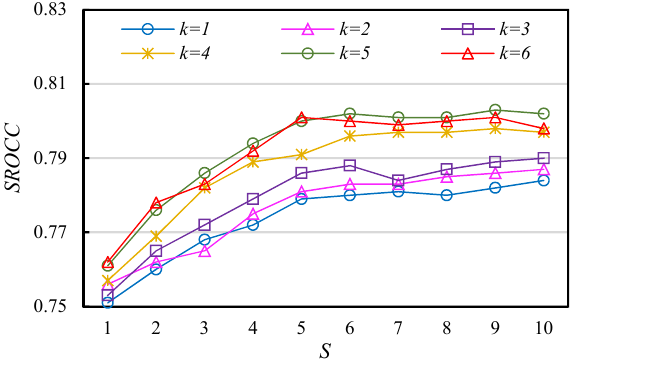}
\caption{The efficacy of parameters $k$ and $S$ in meta-training on LIVE challenge database measured by SROCC.}
\label{fig:2}
\vspace{-0.3cm}
\end{figure}

\subsection{Parameters discussion}
Finally, we conduct experiments to discuss the efficacy of two key parameters in the meta-training of our approach, i.e. $k$ to control the number of NR-IQA tasks in a mini-batch and $S$ to control the learning steps of each task. We set $k$ and $S$ to different values and show the SROCC results on LIVE challenge database in Figure~\ref{fig:2}. The quality evaluation performance of our approach increases with the increase of $k$ and $S$. If $k$ is larger than 5, the SROCC values of our method drop slightly. When $S$ increases from 1 to 6, the performance of quality evaluation increases dramatically. If $S$ is larger than 6, the SROCC values tend to be stable. Therefore, we set $k=5$ and $S=6$ in all the experiments.

\vspace{-0.1cm}
\section{Conclusion}
\label{sec:5}
In this paper, we propose to address the generalization problem of NR-IQA tasks by using meta-learning. We introduce a meta-learning based NR-IQA method with bi-level gradient optimization to learn the shared prior knowledge model of various distortions from a number of NR-IQA tasks, and then fine-tune the prior model on the training data of a NR-IQA task with unknown distortions to obtain the target quality model. Since our model can refine the shared meta-knowledge among various types of distortions when human evaluate image quality, the learned meta-model is easily generalized to unknown distortions. Experiments conducted on five public IQA databases have demonstrated that our approach is superior to the state-of-the-art NR-IQA methods in terms of both generalization ability and evaluation accuracy. In addition, the quality prior model learned from synthetic distortions can also be quickly adapted to the quality assessment of authentically distorted images, which also sheds light on the design of quality evaluation models for real-world applications. \\
\textbf{Acknowledgements.} This work was supported by the National Natural Science Foundation of China (61771473, 61991451 and 61379143), Natural Science Foundation of Jiangsu Province (BK20181354), the Six Talent Peaks High-level Talents in Jiangsu Province (XYDXX-063).

{\small
\bibliographystyle{ieee_fullname}
\bibliography{References}
}

\end{document}